\documentstyle[hx37_latex]{article}

\def\be{\begin{equation}}
\def\ee{\end{equation}}
\def\bea{\begin{eqnarray}}
\def\eea{\end{eqnarray}}

\begin{document}

\title{THE FUTURE OF HST AND THE PLANNING FOR A NEXT GENERATION SPACE TELESCOPE}

\author{R. A. E. FOSBURY~\footnote{Affiliated to the Astrophysics Division,
Space Science Department, European Space Agency}}

\address{Space Telescope - European Coordinating Facility,\\
Karl-Schwarzschild-Str.~2, D-85748 Garching bei M\"unchen, Germany}


\maketitle\abstracts{
After the success of the refurbishment mission to HST in December
1993, the telescope has taken its place at the very leading edge of
astrophysical research. The breadth and depth of the topics covered at the
HST conference in Paris in December 1995 gave us a foretaste of the value
of the data archive being accumulated and the supreme importance of future
observations for guiding and enriching the programmes carried out with Keck
and the very large groundbased telescopes nearing completion. Here I 
outline, as far as we know them, the plans for HST beyond the 1997 and
1999 servicing missions. The current planning for a Next Generation Space
Telescope as part of NASA's {\it Origins} initiative is described with
particular emphasis on possibilities for European participation.}

\section{Introduction}

The question of the future of ultraviolet through infrared astronomy from space
has recently been reviewed by the `HST and Beyond' committee chaired by  Alan
Dressler of the Carnegie Observatories. Their report entitled  `Exploration and
the search for origins:  A vision for ultraviolet-optical-infrared space
astronomy', available from: 

\verb+http://saturn1.gsfc.nasa.gov/ngst/Background/HST_Beyond.PDF+
 
\noindent 
follows and complements the report of the `Toward other planetary systems'
committee  chaired by Bernie Burke of NASA's Solar System Exploration Division.

The principal recommendations are:

\begin{enumerate}

\item{The operation of HST beyond its currently scheduled termination 
date of 2005 using a much lower-cost operation.}

\item{The development of a space observatory of 4m or larger aperture,
optimised for imaging and spectroscopy over the 1--5~$\mu$m\ wavelength range.}

\item{The development of the capability for space interferometry.}

\end{enumerate}

In addition, the committee noted how important it is for scientists to explain 
their motivations, goals and results to the public at large. It is in this spirit that
these activities fall under the umbrella of NASA's {\it Origins} initiative.

This article sketches briefly the schedule for future HST servicing missions
and then outlines the plans which are currently being made in the US for a Next
Generation Space Telescope (NGST) with an aperture approaching 8m.

\section{The future of HST}

The future plans for HST servicing missions, currently planned for 1997, 1999
and  2002, are described by Chris Blades in these proceedings and details are
given on the STScI Web pages at:
 
\verb+http://www.stsci.edu/servmiss/servmiss.html+

\noindent
The  two new instruments to be installed during the second mission (STS-82,
February 1997) are described here by David Axon and Rodger Thompson (Near
Infrared Camera  and MultiObject Spectrograph) and by Bruce Woodgate (Space
Telescope Imaging  Spectrograph).

In 1999, there will be a major reboost of the spacecraft to counteract the
orbital  decay induced by the Solar maximum. New solar arrays will be fitted and
the ESA Faint  Object Camera will be replaced by the Hopkins/Ball Advanced Camera
for Surveys  which is described here by Garth Illingworth.

In 2002, there is an opportunity for one or two new instruments to be installed.
NASA are  about to issue an announcement of opportunity with a due date for
proposals several months  after the second servicing mission. There is clear
potential for international collaboration in the building of at least one of
these instruments.

The plans beyond 2005 are currently uncertain and the possibilities for an
extended, low-cost  operation are being actively investigated. It is clear that
the transition between the scientific operations of HST and NGST will have to be
carefully managed within the budgetary constraints. It must be remembered,
however, that the NGST will be optimised  for longer wavelenghts than HST ---
which will remain the only major observatory for ultraviolet astronomy.

\section{The planning for NGST}

Following the first meeting about NGST, at STScI in December 1995, the study has
gained considerable momentum and so this section is necessarily a snapshot of the
developments which will lead to an interim report in November 1996 and a final
report in mid-1997. Up-to-date information on the status of the 
various study components is available at:

\verb+http://saturn1.gsfc.nasa.gov/ngst/+
 
\noindent
In particular, the study outline at: 

{\verb+http://saturn1.gsfc.nasa.gov/ngst/top/Studyp/STUDYP~1.html+
 
\noindent
is a compact and interesting reference.

Three independent studies are being carried out and their results will be merged
during  August and September 1996. These are being carried out by:

\begin{itemize}

\item{The Lockheed Martin Corporation}

\item{TRW Civil \& International Systems Division}

\item{The Goddard-led Study Team}

\end{itemize}

Of these, only the deliberations of the latter were accessible at the time 
of this meeting and so only these are summarised here.

\subsection{Scientific drivers}

Four categories of science drivers were considered in shaping the mission 
of which only priorities 1 and 2 are listed here:

\vskip .4cm
\noindent
Priority 1 (central to mission)

\begin{itemize}

\item{See galaxy formation: implies large near-IR telescope optimised for 
1--5~$\mu$m, radiatively cooled to less than 70K, 
zodiacal light sensitivity limited, 60--100~mas 
resolution at 1--2~$\mu$m}

\item{Find high-z supernovae: implies wide field cameras
with FOV greater than 3$\times$3~arcmin with more than 4096$^2$\ pixels}

\item{1.5~yr survey, 3.5~yr GO programme: implies lifetime of 
expendables}

\item{Find first globular clusters: implies collecting area greater 
than 12m$^2$}

\item{Measure many z's of `red' galaxies: implies low 
spectral resolution `3D' multi-object spectrograph}

\end{itemize}

\noindent
Priority 2 (very important)

\begin{itemize}

\item{See star formation (post-ISO/SIRTF): implies 
thermal-IR 5--30~$\mu$m camera/spectrograph}

\item{Study foreground galaxies: implies wide field visible camera}

\item{All-sky pointing once/yr for GOs: defines sunshield shape and need to 
look perpendicular to sun}

\item{Monitor supernovae for 2.5~months: constrains sunshield shape and angle 
with respect to sun}

\item{Follow-up for 3 GO cycles: implies 10~year lifetime goal}

\item{$R=1000$\ spectroscopy at $z=5$\ for 5$\sigma$\ sources: implies
large aperture, 8m goal}

\end{itemize}

\subsection{The working groups}

The Goddard-led study has set up several working groups which meet  independently
but regularly coordinate their activities. These include: an
Engineering/Management Study Group; an Industry Advisory Board and groups
looking at Operations Systems; the Optical Telescope Assembly;  the Science
Instrument Module; the Science Performance; the Spacecraft  Support System and
the NGST Systems. In addition, there is a Science  Advisory Committee which provides
input to all three studies.

The budgetary guidelines set by NASA for the study are for a  spacecraft cost of
\$M~500 (phase C/D) and a total mission cost (to include launch and operations)
of some \$M~900. It is clear from this that the mission has to be very different
in concept  from HST and in particular has to be built quickly using proven
technology. The absence of the need for astronaut access, while greatly reducing 
costs, means a somewhat greater mission risk. The operational costs will also
have to be kept strictly controlled.

\subsection{A spacecraft concept}

At the time of this meeting, the Goddard-led study was concentrating on a
passively  cooled 8m (almost filled) aperture, diffraction limited at 2~$\mu$m.
The orbit would be chosen to minimise environmental disturbances, to allow stable
communication links and stability for the nominal mission lifetime with minimum
observation  interruptions and maximum observational opportunities.  The
trajectory to orbit would  be chosen to keep the launch requirements within the
limits of available launchers  and to keep the transfer time as short as
possible. The most promising solution appears to be an `L2 halo' orbit (in the
Sun/Earth system) at a distance of some 1.5 million km from Earth. Such a
location has enormous advantages over low-earth orbit by removing many observing
constraints.  Communications are easier than with heliocentric orbits and there
is full-sky coverage with opportunities for long continuous observations. The
stable thermal  environment facilitates passive cooling.

The required sun-shield limits the instantaneous pointing of the telescope to a 
ring with an axis (V3) along the spacecraft-sun vector and with a width which is
set by the detailed shield shape but would typically be $+5^\circ/-57^\circ$\ from
the  plane perpendicular (V1-V2) to this vector around 
the V2 axis (nb. V1 is the optical axis of the telescope).

As a general tool for studying the effect of variations of NGST design on the
achievement of the scientific goals (as presented in the `HST and Beyond' report),
the STScI has developed a computer program which computes the fraction
of the science programme achieved in a given mission lifetime as a function 
of telescope aperture. This work is described at:

\verb+http://augusta.stsci.edu/+

\noindent

The current design of the Optical Telescope Assembly (OTA) is for a segmented 8m 
aperture with an f/1.25 primary, an f/24 OTA and a 5$\times$5~arcminute FOV. The 
4-mirror, centred design has small M3 and M4 mirrors located behind the primary.
The primary is re-imaged onto a flat, but deformable quatenary mirror for fine 
figure control. With the fast primary and off-axis field, there is very little 
obscuration. A fine steering mirror --- fed with an error signal from the main
camera --- would provide astrometric pointing, avoiding  the necessity of
providing milliarcsecond {\it spacecraft}\ stability. Various concepts for ultra-thin,  lightweight
mirror panels are being investigated.

The baseline launch vehicle is an Atlas IIAs but it is clear that a larger
launcher would allow sufficient aperture for the science programme with a much
simpler deployment scheme.

\subsection{Instruments}

The science instrument suite currently being studied consists of:

\begin{itemize}

\item{a NIR camera with a 4$\times$4~arcmin FOV using 64 1024$^2$ InSb 
(0.6--5~$\mu$m) arrays in four optical assemblies --- rather like the HST WFPC}

\item{a NIR multi-object spectrograph with a 3$\times$3~arcmin FOV:
this and the NIR camera would be passively cooled to less than 40K}

\item{a thermal-IR camera with a 2$\times$2~arcmin FOV using a 1024$^2$ 
Si:As array}

\item{finally, a thermal-IR spectrograph would feed the same detector 
as the TIR camera: these last two instruments would work from 5--26~$\mu$m
and would need to be actively cooled to about 8K}

\end{itemize}

These four instruments would be packaged together and would use adjacent 
parts of the focal plane.

\subsection{Communications and operations}

The data transmission requirements are important because of the relatively  large
distance from Earth (10s round-trip light travel time). At an estimated  40,000
frames per year (5~Tbyte/yr), the average sustained data rate would be
approximately 160~kbyte/s. A single dedicated 11m aperture ground antenna working
with dual s-band and x-band frequencies would provide 8h/day coverage. The NASA
Deep Space Network would be used only in emergencies.

There are many considerations which determine the operational modes and the
nature of the orbit means that they would be different, and generally simpler,
than for HST. Using common microprocessors throughout and employing a
workstation/network environment would avoid much parallel development of
different systems. The number of available observing modes per
instrument/detector would be limited to about four and there would be easy mode
changes for flexible scheduling. One of the major simplifications compared with
the current HST operation would be the use of an autonomous guide star
acquisition using a sub-array of the NIR camera. This avoids the need for costly
pre-selection of guide stars but may result in the inability to make repeated
pointings to exactly the same part of the sky.

This opportunistic pointing control would allow an adaptive scheduling scheme
which means that, after observing programme design, there need be relatively
little intervention from the ground except for error/problem resolution.

\section{European involvement}

The NGST feasibility study in the US is proceeding very fast and, judging by the
excellent quality of the reports and presentations, it is a serious effort with
participants clearly having confidence that the project will proceed. An obvious
limitation of the current concept is the small size of the baseline launcher (the
Atlas IIAs). A larger launcher would make NGST simpler and cheaper --- although
not necessarily larger. Since the low cost of the mission will demand a very
rapid construction schedule, it is clear that all the required technologies will
need to be in place before the building phase of the project.  There will be a number of 
technological studies which need to be carried out and these may be in common
with those needed for different components of the {\it Origins}\ and other 
programmes. It is clear also that there will be opportunities for participation
in the operational phase.

Given the very large investment in groundbased facilities by Europe, a substantial
involvement both in the future of HST (beyond 2001) and in the NGST is vital for
the health and vigour of European astronomy. Such facilities are so mutually 
complementary that they enormously amplify each others value.

\section*{Acknowledgments}
I should like to acknowledge the warmth and openness of colleagues  at NASA, 
the STScI and other US organisations who have welcomed --- without any formal
guarantees of contribution --- the participation
of ST-ECF staff in these studies.

The viewgraphs prepared for this talk are available on the WWW at:

\verb+http://ecf.hq.eso.org/~rfosbury/Beyond_HST/+

\end{document}